\documentclass[12pt]{article}

\usepackage{amsmath}

\usepackage{amssymb}
\usepackage{epsfig}

\newcommand{\ope}{operator product expansion}
\newcommand{\pd}{\partial}

\newcommand{\np}[3]{\big[ #1, #2 \big]_{#3}}

\newcommand{\sub}[2]{{#1}_{\scriptscriptstyle #2}}

\newcommand{\phase}{\mathrm{e}^{\mathrm{i}\pi/4}}
\newcommand{\phaseab}
{\mathrm{e}^{\frac{\mathrm{i}\pi}{4} \sigma_{\alpha \beta} }}
\newcommand{\cpl}{g}
\newcommand{\CC}{\mathbf{C}}

\newcommand{\OO}[1]{O\big( (z-w)^{#1} \big)}
\newcommand{\pole}[2]{\frac{#2}{(z-w)^{#1}}}
\newcommand{\firstpole}[1]{\frac{#1}{z-w}}

\newcommand{\ket}[1]{{|}#1{>}}
\newcommand{\zz}{{\mathbb{Z}_2 \times \mathbb{Z}_2}}

\begin{document}

\begin{titlepage}

\phantom{a}

\vfill

\begin{center}
{\LARGE \bf
$\zz$ graded superconformal algebra
 of parafermionic type
 \\
}
\vspace{20pt}
{
{ \large \bf Boris Noyvert
\\} 
\vspace{10pt}
{  e-mail:
\hspace{-15pt}
\raisebox{-0.7pt}{
\includegraphics[height=8pt]{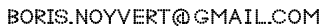}
}\\}
\vspace{3pt}
{ \small \it Department of Mathematics,}\\
{ \small \it  University of York,} \\
{ \small \it  YO10 5DD, York,  United Kingdom.}\\
}

\vspace{20pt}

\begin{abstract}

{\normalsize
We present a new conformal algebra. It is
$\zz$ graded and generated by three $N=1$
superconformal algebras coupled to each other
by nontrivial relations of parafermionic type.
The representation theory and unitary models of the algebra
are briefly discussed. We also conjecture the existence
of infinite series of parafermionic algebras containing
many $N=1$ or $N=2$ superconformal subalgebras.
}

\end{abstract}

\end{center}

\vfill

\end{titlepage}

\tableofcontents


\section{Introduction}

\label{Introduction}


Superconformal algebras are of great importance in theoretical
physics. Probably the best known ones 
are the $N=1$ and the $N=2$ superconformal algebras,
which play an important role in superstring theory.
Here the $N$ is the number of supersymmetry generators.
The $N=1$ algebra was introduced in~\cite{Ramond:1971gb}
and~\cite{Neveu:1971rx}, the $N=2$ algebra first appeared
in~\cite{Ademollo:1975an}. The mathematical meaning of
the term ``superalgebra'' is that the algebra is $\mathbb{Z}_2$
graded. There are even (bosonic) generators and odd (fermionic)
generators. The algebraic relations respect the $\mathbb{Z}_2$
grading.

In this paper we introduce a new algebra,
the superconformal algebra graded by the $\zz$ group.
The $\zz$ group is a finite abelian group containing
4 elements: the identity $(0,0)$, and 3 more
elements $(1,0), (0,1), (1,1)$. The product
of two different non-identity elements gives the third one.
The square of a non-identity element gives identity,
hence there are 3 different $\mathbb{Z}_2$ subgroups in $\zz$.

We take one superconformal generator field
of conformal dimension $3/2$
for each non-identity
element of $\zz$: $G^{(\alpha)}, \alpha=1,2,3$. Each one of them
generates the standard $N=1$ superconformal algebra:
\begin{align}
                      \label{N=1 ope 1}
G^{(\alpha)}(z)G^{(\alpha)}(w)&=
\pole{3}{1}
+\firstpole{\frac{3}{c}T^{(\alpha)}(w)}+
\OO{0},\\
                      \label{N=1 ope 2}
T^{(\alpha)}(z)G^{(\alpha)}(w)&=
\pole{2}{\frac{3}{2}G^{(\alpha)}(w)}
+\firstpole{\pd G^{(\alpha)}(w)}+
\OO{0},\\
                      \label{N=1 ope 3}
T^{(\alpha)}(z)T^{(\alpha)}(w)&=
\pole{4}{c/2}+
\pole{2}{2T^{(\alpha)}(w)}+
\firstpole{\pd T^{(\alpha)}(w)}+
\OO{0}.
\end{align}
The Virasoro fields $T^{(\alpha)}(z), \alpha=1,2,3$
belong to the $(0,0)$ grading.
The \ope\ of two different superconformal
generators should give the third one:
\begin{equation}
G^{(\alpha)}(z) G^{(\beta)}(w)
\sim
\pole{3/2}{G^{(\gamma)}(w)},
\qquad
\alpha \ne \beta \ne \gamma.
\end{equation}
The power of the singularity (3/2) is obtained by a simple
dimensional analysis. The crucial point is that it is not integer.
So our algebra is not a standard chiral algebra
(vertex algebra in mathematical literature),
but a parafermionic type algebra (generalized
vertex algebra). The full algebra as we show in this paper
is formed by 10 generating fields.
In addition to the 6 generating fields
$G^{(\alpha)}(z), T^{(\alpha)}(z),\ \alpha=1,2,3$ mentioned above
one has 3 dimension-$5/2$ fields $U^{(\alpha)}(z),\ \alpha=1,2,3$
and one dimension-3 field $W(z)$.
We call this algebra ``the $\zz$ graded $N=1$ superconformal algebra''.

The first example of parafermionic algebra was introduced
by Fateev and Zamolodchikov
in~\cite{FZ:parafermions:Zn}. This $\mathbb{Z}_N$
graded algebra is generated by $N-1$ fields of conformal dimensions
$\Delta_i=i(N-i)/N,\ i=1,2,\ldots,N-1$. For a fixed $N$ the algebra
has no free parameters. In their next paper~\cite{FZ:parafermions:4/3}
the same authors
presented another $\mathbb{Z}_3$ graded parafermionic algebra,
generated by the Virasoro field and two dimension-$4/3$ fields.
This algebra has a continuous free parameter - the central charge.
Later Gepner~\cite{Gepner:1987sm} introduced new parafermionic
theories through coset construction of the type ${\mathfrak{g}}_k/u(1)^r$,
where $\mathfrak{g}_k$ is the affine Lie algebra on level $k$ and
$r$ is its rank.
The mathematical treatment of parafermionic algebras was developed
in~\cite{Dong_Lepowsky} (see also the recent paper~\cite{Bakalov_Kac}).

In our previous work~\cite{Noyvert:2006qp} we applied the
algebraic approach to calculate the structure constants
of the $sl(n)_2/u(1)^2$ and the $sl(2|1)_2/u(1)^2$ coset parafermions.
We called the generators of the former theory the $sl(n)$ fermions.
The $sl(3)_2/u(1)^2$ and the $sl(2|1)_2/u(1)^2$ parafermionic
algebras are also $\zz$ graded.

The current paper is the direct continuation of~\cite{Noyvert:2006qp},
we use the same setting and the same tools to derive the
$\zz$ graded $N=1$ superconformal algebra. This algebra
resembles in many aspects the $sl(3)$ fermion algebra from~\cite{Noyvert:2006qp}. But the new algebra
is more complicated: it has more generating
fields and has one free continuous parameter.

The paper is organized as following.
First in Section~\ref{Parafermionic conformal algebras}
we recall the main points of the algebraic approach
to conformal algebras of parafermionic type.
In Section~\ref{Algebra} the
$\zz$ graded $N=1$ superconformal algebra is derived.
The full set of lengthy \ope s defining the algebra
is listed in Appendix~\ref{OPEs}.
In Section~\ref{Generalized commutation relations}
we convert the \ope s to the generalized commutation relations
between the modes of the basic fields, preparing the
ground to the study of representation theory of the algebra
(in Section~\ref{Representation theory}).
The unitarity restrictions are discussed in Section~\ref{Unitary models}.
Two explicit realizations of unitary models possessing the
$\zz$ graded $N=1$ superconformal symmetry are presented
in Section~\ref{Explicit realizations}. The last Section~(\ref{Discussion})
contains a brief summary and the ideas for the further study.
In particular we announce the $N=2$ superconformal analogue of
the algebra described in this paper and also announce the existence
of two series of more complicated
$N=1$ and $N=2$ superconformal algebras of parafermionic type and
speculate about their unitary minimal models.



\section{Parafermionic conformal algebras}

\label{Parafermionic conformal algebras}


In this section we briefly recall
the main points
of the algebraic formalism for parafermionic conformal
algebras.
We will follow here Ref.~\cite{Noyvert:2006qp}
(Sections 2 and 3).
This algebraic approach in fact goes back
to 1993~\cite{Dong_Lepowsky}.
See also the recent
paper~\cite{Bakalov_Kac}, where parafermionic algebras
are defined using the notion of polylocal fields.

An \ope\ of parafermionic type has the following form:
\begin{equation}
                                      \label{ope}
A(z) B(w) =\frac{1}{(z-w)^{\alpha}}
\Big(
\np{A}{B}{\alpha}(w)+\np{A}{B}{\alpha-1}(w)(z-w)
+\np{A}{B}{\alpha-2}(w)(z-w)^2+\cdots
\Big),
\end{equation}
i.e.~it is a general \ope\ with one important
restriction that
except the overall singularity $(z-w)^{-\alpha}$
the integer powers of $(z-w)$ only
are present on the right hand side of the
equation. But the singularity $\alpha$ doesn't have to be integer!
Here we also introduced a notation $\np{A}{B}{n}$,
the $n$-product of fields $A$ and $B$. It is the field,
arising at the $(z-w)^{-n}$ term of the \ope\
of the fields $A(z)$ and $B(w)$ around $w$
as it appears in (\ref{ope}).

When $\alpha\notin \mathbb{Z}$ it is not clear a priori
how to exchange the fields in the \ope , since some phases
are involved.
The following axiom, which is
the crucial point of the definition of parafermionic algebras,
tells us how to exchange the fields in the \ope\ (\ref{ope}):
\begin{equation}
                             \label{mutual locality}
A(z)B(w)(z-w)^\alpha=
\mu_{\scriptscriptstyle A B}
B(w)A(z)(w-z)^\alpha.
\end{equation}
Here $\mu_{\scriptscriptstyle A B}$ is a {\it commutation factor}
 which is a complex number
different from zero. 
The exponent $\alpha$ in~(\ref{mutual locality})
is usually chosen to be equal to the
singularity of the \ope. However one can add to $\alpha$ an integer number.
If the integer is even, then the commutation factor is not
changed, if we shift $\alpha$ by odd integer then the
sign of the commutation factor is flipped.

By exchanging the fields in (\ref{mutual locality}) second time
one shows that the commutation factor should satisfy the following
consistency conditions:
\begin{equation}
\mu_{\scriptscriptstyle A B}
\mu_{\scriptscriptstyle B A}=1,
\end{equation}
and if we assume that the term
$\np{A}{A}{\alpha_{\scriptscriptstyle A A}} \ne 0$
then it follows that
\begin{equation}
\mu_{\scriptscriptstyle A A}=1.
\end{equation}

If the \ope\ of two basic fields $B(w)$ and $C(v)$
gives a third one $D(v)$:
\begin{equation}
                          \label{B(w)C(v)}
B(w)C(v)=\frac{D(v)}{(w-v)^{\alpha_{\scriptscriptstyle B C}}}+ \cdots,
\end{equation}
then exchanging another basic field $A(z)$
with $B(w)$ and then with $C(v)$ is essentially the same
as exchanging $A(z)$ with $D(v)$. Therefore
$\sub{\mu}{A D}$ is proportional to $\sub{\mu}{A B} \sub{\mu}{A C}$:
\begin{equation}
                                    \label{mumu}
\sub{\mu}{A B}\, \sub{\mu}{A C} =  \sub{\mu}{A D}
(-1)^{\sub{\alpha}{A B}+\sub{\alpha}{A C}-\sub{\alpha}{A D}}.
\end{equation}
It is also implicitly stated here that
$\sub{\alpha}{A B}+\sub{\alpha}{A C}-\sub{\alpha}{A D}\in \mathbb{Z}$.

The most important tool in the study of parafermionic
conformal algebras is the generalized Jacobi identities. This identities
involve the \ope s between three fields:
\begin{equation}
                                      \label{Jacobi}
\begin{aligned}
&\sum_{j \ge 0}
(-1)^j {\gamma_{\scriptscriptstyle A B} \choose j}
\np{A}{\np{B}{C}{\gamma_{\scriptscriptstyle B C}+1+j}}
{\gamma_{\scriptscriptstyle A B}+\gamma_{\scriptscriptstyle A C}+1-j}
\\
-
\mu_{\scriptscriptstyle A B}
(-1)^{\alpha_{\scriptscriptstyle A B}-\gamma_{\scriptscriptstyle A B}}
&\sum_{j\ge 0}
(-1)^j {\gamma_{\scriptscriptstyle A B} \choose j}
\np{B}{\np{A}{C}{\gamma_{\scriptscriptstyle A C}+1+j}}
{\gamma_{\scriptscriptstyle A B}+\gamma_{\scriptscriptstyle B C}+1-j}\\
=&
\sum_{j\ge 0}
 {\gamma_{\scriptscriptstyle A C} \choose j}
\np{\np{A}{B}{\gamma_{\scriptscriptstyle A B}+1+j}}{C}
{\gamma_{\scriptscriptstyle B C}+\gamma_{\scriptscriptstyle A C}+1-j}\, ,
\end{aligned}
\end{equation}
The sums are finite, the upper bound is given
by the order of singularity of the corresponding fields.
The parameters $\gamma$ differ from the corresponding singularity
exponents $\alpha$ by an integer number:
$\sub{\alpha}{A B}-\sub{\gamma}{A B},
\sub{\alpha}{A C}-\sub{\gamma}{A C},
\sub{\alpha}{B C}-\sub{\gamma}{B C} \in \mathbb{Z}$.


\section{Derivation of the algebra}

\label{Algebra}


As we have already mentioned in the introduction
we start from three copies of the $N=1$ superconformal algebra,
associated to the three 
non-identity elements of the $\zz$ abelian group.
The generators are $G^{(\alpha)}(z),T^{(\alpha)}(z),\ \alpha=1,2,3$.
The algebraic relations inside the $N=1$ superconformal algebra
are given by the \ope s~(\ref{N=1 ope 1},\ref{N=1 ope 2},\ref{N=1 ope 3}).
Note the unusual normalization of
superconformal generators $G^{(\alpha)}(z)$.
The parameter $c$ is the central charge
of the three $N=1$ superconformal algebras.

Now we want to couple the fields $G^{(\alpha)}(z)$
to each other. The \ope\ of two superconformal generators $G$
gives the third one:
\begin{align}
G^{(\alpha)}(z)G^{(\beta)}(w)&=
\pole{3/2}{\kappa_{\alpha, \beta}G^{(\gamma)}(w)}+
\OO{-1/2},
\end{align}
where $\kappa_{\alpha, \beta}$ are yet unknown structure constants and
$\alpha, \beta, \gamma$ are all different.

The fields in these \ope s are exchanged as following:
\begin{equation}
G^{(\alpha)}(z)G^{(\beta)}(w)(z-w)^{3/2}=
\mu_{\alpha,\beta}G^{(\beta)}(w)G^{(\alpha)}(z)(w-z)^{3/2},
\end{equation}
leading to the relations between the structure constants
and the commutation factors:
\begin{equation}
\kappa_{\alpha, \beta}=
\mu_{\alpha,\beta}\kappa_{\beta, \alpha }.
\end{equation}

The commutation factors are easily determined using
the relation (\ref{mumu}) between them.
Taking $A=G^{(1)}, B=G^{(1)}, C=G^{(2)}$
we get
\begin{equation}
\mu_{1, 1}\mu_{1, 2}=-\mu_{1, 3},
\end{equation}
since the singularities are equal to
$\alpha_{1, 1}=3, \alpha_{1, 2}=\alpha_{1, 3}=3/2$,
and so \\
 $(-1)^{\alpha_{1, 1}+\alpha_{1, 2}-\alpha_{1, 3}}=-1$.
Substituting $A=G^{(1)}, B=G^{(2)}, C=G^{(3)}$ in (\ref{mumu})
we get
\begin{equation}
\mu_{1, 2}\mu_{1, 3}=\mu_{1, 1}.
\end{equation}
Taking into account that $\mu_{1, 1}=1$, one obtains
\begin{equation}
\mu_{1, 2}=-\mu_{1, 3}=\pm \mathrm{i}.
\end{equation}
To resolve the formal ambiguity we fix $\mu_{1, 2}= \mathrm{i}$.
Using the cyclic permutations of indices we determine all the
commutation factors:
\begin{equation}
                      \label{mu=i}
\mu_{1, 2}=\mu_{2, 3}=\mu_{3,1}=-\mu_{2,1}
=-\mu_{3,2}=-\mu_{1,3}= \mathrm{i}.
\end{equation}

To determine the structure constants we use
the generalized Jacobi identities~(\ref{Jacobi}).
Take $A=G^{(1)}, B=G^{(2)}, C=G^{(3)}$
and two parameters from the set of three
$\sub{\gamma}{A B}, \sub{\gamma}{B C}, \sub{\gamma}{A C}$
equal to $1/2$ and the third one
equal to $3/2$. Then the corresponding Jacobi identities
require that the structure constants are equal to each other:
\begin{equation}
\kappa_{1, 2}=\kappa_{2, 3}=\kappa_{3,1}=\cpl\, \phase,
\end{equation}
where we introduced the new phase shifted structure constant $\cpl$
in order to avoid appearance of $\mathrm{i}$ in the formulas below.

We will assign the following
$\mathbb{Z}_2 \times \mathbb{Z}_2$ charges to the fields:
$G^{(1)}$ has charge $(1,0)$, $G^{(2)}$ -  $(0,1)$, $G^{(3)}$ - $(1,1)$.
Then the identity field and all the Virasoro generators $T^{(\alpha)}$
carry the charge $(0,0)$. The commutation factors of the fields
from the $(0,0)$ sector with all the fields are equal to 1,
and the commutation factors between other sectors
are given by~(\ref{mu=i}).

Now it is easy to derive the leading terms in the \ope s
of different generating fields using the dimensional
and $\mathbb{Z}_2 \times \mathbb{Z}_2$ charge analysis:
\begin{align}
T^{(\alpha)}(z)G^{(\beta)}(w)&=
\kappa\left(T^{(\alpha)},G^{(\beta)}\right)
\pole{2}{G^{(\beta)}(w)}+
\OO{-1},
\\
T^{(\alpha)}(z)T^{(\beta)}(w)&=
\pole{4}{\kappa\left(T^{(\alpha)},T^{(\beta)}\right)}+
\OO{-2},
\end{align}
where $\alpha \ne \beta$ and
$\kappa\left(T^{(\alpha)},G^{(\beta)}\right),
\kappa\left(T^{(\alpha)},T^{(\beta)}\right)$
are structure constants to be determined
by the Jacobi identities in the following way.
Insert $A=G^{(\alpha)}, B=G^{(\alpha)}, C=G^{(\beta)}$
($\alpha \ne \beta$) and
$\sub{\gamma}{B C}=\sub{\gamma}{A C}=1/2$,
$\sub{\gamma}{A B}=0$ to the Jacobi identities~(\ref{Jacobi})
to get
\begin{equation}
\kappa\left(T^{(\alpha)},G^{(\beta)}\right)=
(1+16\cpl^2)\, c/24.
\end{equation}
Choose $A=T^{(\alpha)}, B=G^{(\beta)}, C=G^{(\beta)}$
($\alpha \ne \beta$) and
$\sub{\gamma}{B C}=0, \sub{\gamma}{A C}=2, \sub{\gamma}{A B}=1$,
the Jacobi identity then enforces
\begin{equation}
\kappa\left(T^{(\alpha)},T^{(\beta)}\right)=
\frac{c}{3}\kappa\left(T^{(\alpha)},G^{(\beta)}\right)=
(1+16\cpl^2)\, c^2/72.
\end{equation}

All the Jacobi identities for the fields
$G^{(\alpha)}, T^{(\alpha)}, \alpha=1,2,3$,
taking into account only the terms specified in the
above \ope\ relations, are satisfied now.
So this parafermionic algebra is selfconsistent,
there are two free parameters: $c$ and $\cpl$.
However we haven't specified all the singular terms in the
\ope s, so the information contained
in the generalized commutation relations
extracted from the above \ope s is not sufficient to build
the representation theory of the algebra.
We have to specify all the singular terms in the
\ope s of generating fields in terms of the generating fields,
their derivatives and composite fields.
By singular terms we understand all the $n$-products
$\np{A}{B}{n},\ n>0$, by composite field we mean
$\np{A}{B}{n},\ n\le 0$, where $A$ and $B$ are two
generating fields.

We have to make the additional assumptions about the missing
singular terms in the \ope s. The first assumption
is that there are no other dimension-2 fields in the algebra.
It means that the field in $\np{T^{(\alpha)}}{T^{(\beta)}}{2}$
is a linear combination of $T^{(1)}$, $T^{(2)}$ and $T^{(3)}$.
The second consequence of this assumption is that
the total energy-momentum field $T(z)$ is proportional to the
sum of $T^{(1)}(z)$, $T^{(2)}(z)$ and $T^{(3)}(z)$.
The factor is easily calculated from the requirement that
the weight of the fields $G^{(\alpha)}$ under the action of $T(z)$
is equal to their conformal dimension $3/2$. So we get that the total
energy-momentum field is
\begin{equation}
                                      \label{T total}
T=\frac{1}{1+\frac{c}{18}\left(
1+16\cpl^2
\right)}
\left(T^{(1)}+T^{(2)}+T^{(3)}\right).
\end{equation}
The requirement $\np{T}{T^{(\alpha)}}{2}=2T^{(\alpha)}$
leads to fixing the coefficients in the second order term
in the \ope\ of $T^{(\alpha)}$ and $T^{(\beta)}$:
\begin{equation}
\np{T^{(1)}}{T^{(2)}}{2}
=\frac{c}{18}\left(
1+16\cpl^2
\right)
\left(T^{(1)}+T^{(2)}-T^{(3)}\right),
\end{equation}
and the same for cyclic permutations of the indices.

Now it is easy to verify that the energy-momentum
field $T(z)$ indeed satisfies the Virasoro algebra,
the central charge of which is
\begin{equation}
                             \label{total c}
\CC=\frac{3c}{1+\frac{c}{18}\left(
1+16\cpl^2
\right)}.
\end{equation}

Now we look at the next to leading term
$\np{G^{(\alpha)}}{G^{(\beta)}}{1/2}$
in the expansion of two superconformal generators.
This is a dimension-$5/2$ field.
The simplest assumption that it is just proportional
to $\pd G^{(\gamma)}$ does not work. We have to introduce
three new basic fields $U^{(1)}, U^{(2)}, U^{(3)}$ of dimension $5/2$:
\begin{align}
                      \label{GG ope long}
G^{(\alpha)}(z)G^{(\beta)}(w)&=
\phase \cpl
\left(
\pole{3/2}{G^{(\gamma)}(w)}
+\pole{1/2}{\frac{1}{2}\pd G^{(\gamma)}(w)+
U^{(\gamma)}(w)}
\right)+
\OO{1/2},
\end{align}
where $\alpha, \beta, \gamma$ are cyclically ordered.
The coefficient $\frac{1}{2}$ before $\pd G^{(\gamma)}(w)$
is chosen so to make the field $U^{(\gamma)}(w)$ primary
with respect to the total energy-momentum Virasoro field.
The field $U^{(\gamma)}$ has the same $\mathbb{Z}_2 \times \mathbb{Z}_2$
grading as $G^{(\gamma)}$. Consequently $U^{(\gamma)}$ has the same commutation
factors with other fields as $G^{(\gamma)}$, if defined appropriately:
\begin{equation}
\sub{\mu}{A,U^{(\gamma)}}=\sub{\mu}{A,G^{(\gamma)}},
\end{equation}
if $\sub{\alpha}{A,U^{(\gamma)}}=\sub{\alpha}{A,G^{(\gamma)}}
+2\,\mathbb{Z}$
in~(\ref{mutual locality}).
Here $A$ stands for any field.

The first order singular term in the expansion of $T^{(\alpha)}$
and $G^{(\beta)}$ is also of dimension $5/2$ and a priori is not
expressed in terms of $U^{(\gamma)}$ and $\pd G^{(\gamma)}$ only.
But we would like to make the life easy assuming that
no new dimension-$5/2$ basic fields have to be introduced.
This assumption will cost us one free parameter in the algebra:
$c$ becomes a function of $\cpl$.
First note that with respect to the $N=1$ superconformal
algebra generated by $T^{(1)}$ and $G^{(1)}$ the two other
dimension-$3/2$ fields $G^{(2)}$ and $G^{(3)}$ are the highest weight
primary fields of Ramond type of weight $(1+16\cpl^2)\, c/24$:
\begin{equation}
\begin{aligned}
G^{(1)}_0 \ket{G^{(2)}}&=\cpl \, \phase \ket{G^{(3)}}, \quad
&G^{(1)}_n \ket{G^{(2,3)}}&=0,\ n>0,\\
T^{(1)}_0 \ket{G^{(2,3)}}&=\frac{c}{24}(1+16\cpl^2)\ket{G^{(2,3)}}, \quad
&T^{(1)}_n \ket{G^{(2,3)}}&=0,\ n>0.
\end{aligned}
\end{equation}
Then the field $U^{(2)}$ is expressed in terms of
$G^{(1)}_{-1} G^{(1)}_0 \ket{G^{(2)}}$ and $\np{T^{(1)}}{G^{(2)}}{1}$
is expressed in terms of $T^{(1)}_{-1} \ket{G^{(2)}}$.
$G^{(1)}_{-1} G^{(1)}_0 \ket{G^{(2)}}$ and $T^{(1)}_{-1} \ket{G^{(2)}}$
are in general two independent states in the highest weight
representation. However if there is a null state on level 1
in the highest weight representation then
$G^{(1)}_{-1} G^{(1)}_0 \ket{G^{(2)}} \sim T^{(1)}_{-1} \ket{G^{(2)}}$
and $\np{T^{(1)}}{G^{(2)}}{1}$ is expressed through $U^{(2)}$.
The null state appears on level 1 when the highest weight $h$ and
the central charge $c$ are connected by the following equation:
$3 c - 72 h + 16 c h + 128 h^2=0$. Upon substitution
$h=(1+16\cpl^2)c/24$ the relation is translated to
\begin{equation}
                               \label{c_cpl}
c=\frac{54 \cpl^2}{(1+4 \cpl^2)(1+16\cpl^2)}\, .
\end{equation}

So at last we can fix the following \ope s:
\begin{align}
T^{(\alpha)}(z)G^{(\beta)}(w)&=
\frac{c(1+16\cpl^2)}{24}
\left(
\pole{2}{G^{(\beta)}(w)}
+\firstpole{\frac{2}{3}\pd G^{(\beta)}(w)-
\sigma_{\alpha \beta}\frac{4}{3}U^{(\beta)}(w)}
\right)+
\OO{0},
\end{align}
where
$\alpha \ne \beta$
and we introduced the following two-index
symbol:
\begin{equation}
\sigma_{\alpha \beta}=\left\{
\begin{array}{ll}
0, & \alpha = \beta,\\
\epsilon_{\alpha \beta \gamma}, & \alpha \ne \beta \ne \gamma,
\end{array}
\right.
\qquad
\alpha, \beta, \gamma = 1,2,3,
\end{equation}
i.e.~$\sigma_{12}=\sigma_{23}=\sigma_{31}=1$
and $\sigma_{21}=\sigma_{32}=\sigma_{13}=-1$.

Another new basic generator field $W$ of dimension 3 appears in the
first order pole $\np{T^{(\alpha)}}{T^{(\beta)}}{1}, \ \alpha\ne \beta$:
\begin{equation}
                                    \label{TT1 term}
\np{T^{(\alpha)}}{T^{(\beta)}}{1}
=\frac{c\left(
1+16\cpl^2
\right)}{36}
\left(
\pd T^{(\alpha)}+\pd T^{(\beta)}-\pd T^{(\gamma)}\right)
+\sigma_{\alpha \beta} W.
\end{equation}
(A priori there are 3 such fields but since
$\np{T}{T^{(\beta)}}{1}=\pd T^{(\beta)}$ for $\beta=1,2,3$ all
the three dimension-3 primary fields are proportional to $W$.)
The field $W$ has the $\zz$ charge equal to $(0,0)$.
So its \ope s with all the other fields contain only integer
powers of $(z-w)$ and it has the following exchange properties:
\begin{equation}
A(z) W(w)=W(w) A(z)
\end{equation}
for any generator field $A(z)$.

It comes out that no other new fields are needed to close the algebra.
The \ope s of the fields $U^{(\alpha)}$ and $W$ with all the other
basic fields are constructed using the dimensional and $\zz$
charge analysis, taking into account the basic fields
($G^{(\alpha)}, T^{(\alpha)}, U^{(\alpha)}, \alpha=1,2,3$ and $W$),
their derivatives and the composite fields
(e.g.~$\np{G^{(1)}}{G^{(2)}}{-1/2}, \np{T^{(2)}}{G^{(3)}}{0},\ldots$).
The structure constants are fixed by a routine check
of the Jacobi identities. In the end we obtain the \ope s
listed in Appendix~\ref{OPEs}. All the Jacobi identities are satisfied
modulo a null field condition:
\begin{equation}
\pd W+\frac{(1 + 16 {\cpl}^2)c^2}{27}
\left(
\np{G^{(1)}}{U^{(1)}}{0}+
\np{G^{(2)}}{U^{(2)}}{0}+
\np{G^{(3)}}{U^{(3)}}{0}
\right)
=0.
\end{equation}

We want also to discuss here the subalgebras of the
$\zz$ graded $N=1$ superconformal algebra. Obviously it has
the three $N=1$ superconformal subalgebras generated
by $G^{(\alpha)}, T^{(\alpha)}$. Their bosonic parts
generated by $T^{(\alpha)}$ only are also subalgebras.
Another Virasoro subalgebra is generated by $T-T^{(\alpha)}$.
It is commutative with both $T^{(\alpha)}$ and $G^{(\alpha)}$,
so its central charge is equal to $\CC-c$. The
$\zz$ graded $N=1$ superconformal algebra has no other
proper subalgebras.


\section{Generalized commutation relations}

\label{Generalized commutation relations}


The mode expansions of the fields are introduced as
\begin{equation}
                                      \label{mode expansion}
A(z)=\sum_{n 
}A_n z^{-n-\Delta(A)}.
\end{equation}
The generalized commutation relations between the field modes
are obtained using the formula
\begin{equation}
                                      \label{commutator formula m+n}
\sum_{j=0}^\infty
{\scriptstyle \alpha-k \choose  \scriptstyle j}
(-1)^j
\Big(
A_{m+\alpha-k-j} B_{n-\alpha+k+j}- 
\mu_{\scriptscriptstyle A B}(-1)^k
B_{n-j} A_{m+j}
\Big)=
\sum_{l=0}^{k-1}
{\scriptstyle m+\Delta_A -1 \choose \scriptstyle k-1-l} C^{(l)}_{m+n}.
\end{equation}
For the derivation see Section 4 in~\cite{Noyvert:2006qp}.
$A_m$ and $B_n$ are the modes of the fields $A(z)$ and $B(z)$,
and $C^{(l)}(z)$ are the terms in the \ope\ of $A(z) B(w)$:
\begin{equation}
                                      \label{ope_C}
A(z) B(w) =\frac{1}{(z-w)^\alpha}
\Big(
C^{(0)}(w)+C^{(1)}(w)(z-w)+C^{(2)}(w)(z-w)^2+\cdots
\Big).
\end{equation}
The commutation factor $\mu_{\scriptscriptstyle A B}$
in~(\ref{commutator formula m+n}) is chosen with respect
to the $\alpha$ in the same formula.
The integer number $k$ in~(\ref{commutator formula m+n})
is equal to the number of terms in the \ope\
which are taken into account. The generalized
commutation relation with smaller $k$ can be obtained
from that with larger $k$. Taking into account all
the singular terms in the \ope\ is sufficient to build
the representation theory. However, in some calculations
one can use the generalized commutation relation for the smaller
number of terms, which is usually a more simple formula.

The singularities $\alpha$ in the \ope s
of the fields in the $(0,0)$ charge
$\zz$ sector ($T^{(\beta)}, \beta=1,2,3$, and $W$)
with all the fields in the algebra are
integer. Therefore the corresponding generalized commutation
relations are just usual commutators.
The \ope s inside the same $\zz$ sector are also of
standard type, so the relations between
$G^{(\beta)}_n, U^{(\beta)}_m$ (for the same $\beta$)
are anticommutation relations.
The only relations which are of parafermionic type are
those between the $G, U$ fields from different $\zz$
sectors. For example:
\begin{equation}
                        \label{GG commutation}
\begin{aligned}
\sum_{j=0}^\infty
{\scriptstyle j-1/2 \choose \scriptstyle j}
\Big(
{\rm{e}}^{-\frac{{\rm{i}}\pi}{4}}
G^{(1)}_{m-1/2-j} G^{(2)}_{n+1/2+j}
&-{\rm{e}}^{\frac{{\rm{i}}\pi}{4}}
G^{(2)}_{n-j} G^{(1)}_{m+j}
\Big)= \\
&=
\cpl \left(
{{\frac{m-n-1/2}{2}}}G^{(3)}_{n+m}+
U^{(3)}_{n+m}
\right),
\end{aligned}
\end{equation}
and the same for the cyclic permutation of the indices.
Because of the space limitations we will not list all the
generalized commutation relations here.
The one which is important for the discussion below
is the following commutation relation:
\begin{equation}
                         \label{TT commutation}
\begin{aligned}
{[}T^{(1)}_m,T^{(2)}_n{]}&=\frac{(1 + 16 {\cpl}^2)c^2 m(m^2-1)}{432}
 \delta_{0,m+n}+ \\
&+\frac{3 {\cpl}^2(m-n)}{2 \left( 1 + 4 {\cpl}^2 \right)}
\left(
T^{(1)}_{m+n}+T^{(2)}_{m+n}-T^{(3)}_{m+n}
\right)
+W_{m+n},
\end{aligned}
\end{equation}
and the same for the cyclic permutation of the indices.

We should also draw your attention that some \ope s include
composite operators, therefore the generalized commutation relations
will include infinite sums of terms quadratic in modes also on the right hand
side of~(\ref{commutator formula m+n}). This happens for example
in the case of the relation obtained from the \ope\
$G^{(\alpha)}(z)U^{(\beta)}(w),\ \alpha \ne \beta$ (\ref{GU ope}).
The mode expansions of composite operators are derived
in Appendix E of \cite{Noyvert:2002mc}. But here we can just
extract
the mode expansions for the composite operators from the same formula~(\ref{commutator formula m+n}).
One should choose $k$ in such a way that the last term in the \ope\
taken into account is the composite operator we are interested in.
Then the ``reversed'' formula is
\begin{equation}
                           \label{composite mode expansion}
\begin{aligned}
&\left(\np{A}{B}{\beta}\right)_{m+n}=
-\sum_{\gamma=\beta+1}^\alpha
{\scriptstyle m+\Delta_A -1 \choose \scriptstyle \gamma-\beta}
\left(\np{A}{B}{\gamma}\right)_{m+n}+\\
&\phantom{A A A}+
\sum_{j=0}^\infty
{\scriptstyle \beta-1 \choose  \scriptstyle j}
(-1)^j
\Big(
A_{m+\beta-1-j} B_{n-\beta+1+j}+
\mu_{\scriptscriptstyle A B}(-1)^{\alpha-\beta}
B_{n-j} A_{m+j}
\Big).
\end{aligned}
\end{equation}
There is some freedom in the choice of $n$ and $m$
(as long as $n+m$ is not affected).
If $m \in -\Delta_A+\mathbb{Z}$ then the freedom
can be used to simplify the formula:
choose $m=-\Delta_A+1$, then the first sum
in~(\ref{composite mode expansion}) vanishes.

We should stress that all the infinite sums in the formulae
above are nicely ordered in the sense that
large positive modes are always from the right,
so when applied to a state in a highest weight module
the sum is truncated and becomes finite. Due to this fact
we can use the generalized commutation relations in the
computations on highest weight modules.
The relation obtained from the \ope\ $A(z)B(w)$
(even if includes formally infinite sums from both sides
of the relation) should be used for exchanging the
modes $A_n$ and $B_m$. Although the calculations could be
very complicated, they are very formal and can be held
by a computer using the software for symbolic computations,
like {\sl Mathematica} (the one we have used).
By exchanging the modes it should be possible to order them,
i.e.~a kind of Poincare--Birkhoff--Witt theorem should hold,
but we do not know even how to choose the Poincare--Birkhoff--Witt basis.


\section{Representation theory}

\label{Representation theory}


Highest weight states are the states which are annihilated
by positive modes of all the basic fields.
Highest weight representations are obtained from the highest weight
state by application of nonpositive field modes to it.
The modes of the fields which belong to the $(0,0)$ charge $\zz$ sector
($T^{(\beta)}, \beta=1,2,3$, and $W$) are always integer.
The modes of the $G^{(\beta)}, U^{(\beta)}$ generators can be
integer or half-integer depending on which state they are applied to.
One can deduce from the generalized commutation relation~(\ref{GG commutation})
that the states in the highest weight module can be of 4 types.
One is of the ``NS-NS-NS'' type, which means that all the
$G^{(\beta)}_n, U^{(\beta)}_m$ ($\beta=1,2,3$) modes applied
to it are half-integer:
$n,m \in \mathbb{Z}+1/2$.
And three other are of ``NS-Ramond-Ramond'' type,
which means that the modes of $G, U$ fields from one sector
should be half-integer, when applied to this state,
and the modes of $G, U$ fields from two other sectors
should be integer. These 4 types of states correspond
to the 4 elements of the $\zz$ group, and the $\zz$ grading
can be extended from the algebra to its representations
in the following way. There will be highest weight states
of 4 types: $\ket{(0,0)}, \ket{(1,0)}, \ket{(0,1)}, \ket{(1,1)}$,
according to their $\zz$ charge. Then the charge of the state
in the highest weight module is the sum (modulo 2) of
charges of the highest weight state and the field modes applied
to it. The states of $(0,0)$ charge are of course of
``NS-NS-NS'' type. And the states of $(1,0), (0,1), (1,1)$
charge are of ``NS-Ramond-Ramond'' type, if the $G,U$ fields
are in the same sector as the state then their modes are half-integer,
if they are from the different sector then their modes are integer.
To illustrate the above rule we give an example of a valid state:
$G^{(3)}_{-3}G^{(1)}_{-5/2}T^{(3)}_{-5}G^{(3)}_{-3/2}
T^{(2)}_{-2}T^{(1)}_{-3}G^{(1)}_{-1}W_{-2}
G^{(2)}_{-1/2}G^{(1)}_{-1/2}\ket{(1,0)}$,
this state has the $\zz$ charge $(0,1)$.

Next we should discuss the zero modes. First we have to choose
Cartan generators,
a commuting set of zero modes. The eigenvalues
of these operators on a highest weight state will be taken
as weights labelling the highest weight state.
It would be desirable to have the zero modes of the 3 Virasoro
fields as Cartan generators, but unfortunately they do not commute,
since according to~(\ref{TT commutation})
\begin{equation}
{[}T^{(1)}_{0},T^{(2)}_{0}{]}=W_0.
\end{equation}
The maximal commuting set consists of 2 operators only:
e.g.~$T^{(3)}_{0}$ and $T^{(1)}_{0}+T^{(2)}_{0}$.
We will label the highest weight representations by
the total conformal weight, the eigenvalue of
the total energy-momentum field~(\ref{T total}),
and the eigenvalue of one of the 3 Virasoro fields,
say $T^{(3)}_{0}$:
\begin{equation}
                          \label{hws}
\begin{aligned}
A_{n}\ket{h,a,q}&=0, \quad n>0, \\
T_{0}\ket{h,a,q}&=h \ket{h,a,q},\\
T^{(3)}_{0}\ket{h,a,q}&=a \ket{h,a,q},
\end{aligned}
\end{equation}
where $q$ denotes the $\zz$ charge of the highest weight state
and $A$ represents any basic field.
There are two more zero modes coming from the $(0,0)$ sector generators.
In the case $q=(0,0)$ there are no other zero modes,
in the case $q=(1,0), (0,1)$, or $(1,1)$ there are 4 more zero
modes coming from the $G,U$ generators.
Some linear combinations of zero modes (with $h,a,q$ dependent
coefficients) will also annihilate the highest weight state in~(\ref{hws}).


\section{Unitary models}

\label{Unitary models}


All the generalized commutation relations are invariant under the
following conjugation:
\begin{equation}
\begin{aligned}
(G^{(\alpha)}_n)^\dag&=G^{(\alpha)}_{-n},
&(U^{(\alpha)}_n)^\dag&=-U^{(\alpha)}_{-n},\\
(T^{(\alpha)}_n)^\dag&=T^{(\alpha)}_{-n},
&(W_n)^\dag&=-W_{-n},
\end{aligned}
\end{equation}
if the algebra parameter $\cpl$ is real.
This conjugation is compatible with the standard conjugation
on the three $N=1$ superconformal subalgebras.
We know that the $N=1$ superconformal algebra has unitary representations
either when the central charge $c\ge 3/2$, or when $c<3/2$
at the following discrete set of values of the central charge:
\begin{equation}
c_p=\frac{3}{2}-\frac{12}{(p-1)(p+1)}, \qquad p=3,4,5,\ldots,
\end{equation}
which correspond to the unitary minimal models of the
$N=1$ superconformal algebra.
From this we can immediately deduce the restrictions on
possible unitary models of the whole algebra.
In our case the $N=1$ subalgebra central charge is connected
to the coupling $\cpl$ by the formula~(\ref{c_cpl}).
For real $\cpl$ we have $c \le 3/2$.
If $c=3/2$ then $\cpl^2=1/8$ and the total central charge
(calculated from (\ref{total c})) is
$\CC=18/5$. If $c=c_p$, then there are two solutions for
 $\cpl^2$ (and consequently for $\CC$). Both
 solutions can be parameterized
by the same formula but with different ranges for the parameter p:
\begin{equation}
                    \label{p series}
\begin{aligned}
\cpl^2_p&=\frac{p+3}{8 \left( p -3\right) },\\
\CC_p&=\frac{18}{5}\left(
1-
\frac{4 \left(  p+11 \right) }{\left(  p +1\right)
\left(5 p-1 \right)}
\right),
\end{aligned}
\end{equation}
where $p=3,4,5,\ldots$ or $p=-3,-4,-5,\ldots$.
In the case $p=3$ the coupling $\cpl$ becomes formally infinite,
but the algebra still makes sense, one has just to redefine
the generators $G^{(\alpha)}$. The $N=1$ subalgebra central charge $c$
and the total central charge $\CC$ both vanish in this case.

In fact the $p=4$ model ($\cpl^2=7/8, \CC=126/95$) is also
excluded from the candidates for the unitary models, since
the central charge of the Virasoro algebra generated by
the field $T-T^{(3)}$ is equal to $126/95-7/10=119/190$,
it is less than 1, but does not belong to the series of
values of the central charge for the Virasoro algebra minimal models.

We should stress that we have no proof that the algebra indeed has unitary
representations except two models for which we know explicit realization
in terms of unitary fields. These realizations are described in the next
section.


\section{Explicit realizations}

\label{Explicit realizations}


\subsection{$sl(3)$ fermions $\times$ affine $so(3)$ on level 4}

\label{fermions times so}


Here we present the construction of the $\zz$ superconformal algebra
at the central charge $\CC=18/5$
in terms of $sl(3)$ fermions and the $so(3)$ affine Kac-Moody algebra
on level 4.
This construction is in a sense a $\zz$ analogue of the realization
of the standard $N=1$ superconformal algebra at the central
charge $c=3/2$ in terms of one free boson and one free fermion.

The $sl(3)$ fermion system is described in detail in~\cite{Noyvert:2006qp}.
We will briefly recall its definition here.
It is also a $\zz$ graded algebra of parafermionic type,
but the conformal dimensions of the main generating fields
are equal to $1/2$ and not to $3/2$ like in the case of
$\zz$ graded $N=1$ superconformal algebra. The algebra is generated
by 3 fermion fields $\psi^{(\alpha)},\, \alpha=1,2,3$.
The \ope\ of each field with itself is the standard free fermion relation:
\begin{equation}
\psi^{(\alpha)}(z) \psi^{(\alpha)}(w)=\frac{1}{z-w}+O(z-w).
\end{equation}
The \ope\ of two different fields gives the third one:
\begin{equation}
\psi^{(\alpha)}(z) \psi^{(\beta)}(w)=
\frac{c_{\alpha, \beta} \psi^{(\gamma)}(w)}{(z-w)^{1/2}}+
O((z-w)^{1/2}),
\quad
\alpha \ne \beta \ne \gamma.
\end{equation}
The fields in the \ope\ are exchanged using our general prescription:
\begin{equation}
\begin{aligned}
\psi^{(\alpha)}(z) \psi^{(\alpha)}(w)&=-\psi^{(\alpha)}(w)\psi^{(\alpha)}(z),\\
\psi^{(\alpha)}(z) \psi^{(\beta)}(w)(z-w)^{1/2}&=
\sub{\mu}{\alpha,\beta}\psi^{(\beta)}(w) \psi^{(\alpha)}(z)(w-z)^{1/2},
\quad
\alpha \ne \beta.
\end{aligned}
\end{equation}
The commutation factors can be obtained exactly in the same way
as the commutation factors of the $\zz$ graded $N=1$ superconformal algebra
(see section~\ref{Algebra}), the result is:
\begin{equation}
\mu_{1,2}=\mu_{2,3}= \mu_{3,1}=-{\rm i}=
-\mu_{2,1}=-\mu_{3,2}= -\mu_{1,3}.
\end{equation}
The structure constants are determined in~\cite{Noyvert:2006qp}
using Jacobi identities:
\begin{equation}
c_{1,2}=c_{2,3}= c_{3,1}=
\frac{\mathrm{e}^{-\frac{\mathrm{i}\pi}{4}}}{\sqrt{2}},
\qquad
c_{2,1}=c_{3,2}= c_{1,3}=
\frac{\mathrm{e}^{\frac{\mathrm{i}\pi}{4}}}{\sqrt{2}}.
\end{equation}
The $sl(3)$ fermion model is given by the following
coset construction \cite{Gepner:1987sm}:
\begin{equation}
\frac{sl(3)_2}{u(1)^2}.
\end{equation}

The second part of our construction is the $so(3)$ affine vertex algebra.
It is also generated by 3 fields, and the algebra is also $\zz$ graded.
The fields $J^{(\alpha)}(z)$ are of conformal dimension $1$, the defining
\ope\ is:
\begin{equation}
J^{(\alpha)}(z)J^{(\beta)}(w)=
\pole{2}{k\, \delta_{\alpha,\beta}}+
\firstpole{\mathrm{i}\epsilon_{\alpha \beta \gamma} J^{(\gamma)}(w)}+
\OO{0}.
\end{equation}

The $sl(3)$ fermions and the affine currents commute:
\begin{equation}
\psi^{(\alpha)}(z)J^{(\beta)}(w)=
J^{(\beta)}(w)\psi^{(\alpha)}(z)=
\OO{0}.
\end{equation}

The superconformal generators $G^{(\alpha)}$ of the
$\zz$ graded $N=1$ superconformal algebra are expressed as products
of corresponding $sl(3)$ fermions and affine currents:
\begin{equation}
G^{(\alpha)}(z)=\frac{1}{\sqrt{k}}\, \psi^{(\alpha)}(z)J^{(\alpha)}(z).
\end{equation}
Then the $\cpl$-coupling of the $\zz$ graded $N=1$ superconformal algebra
is equal $\cpl=1/\sqrt{2k}$.

The Virasoro field associated with $G^{(\alpha)}(z)$ is
\begin{equation}
T^{(\alpha)}=\frac{1}{2k}\, \np{J^{(\alpha)}}{J^{(\alpha)}}{0}+
\frac{1}{2}\,\np{\psi^{(\alpha)}}{\psi^{(\alpha)}}{-1}=
\frac{1}{2k}\, {:}{J^{(\alpha)}}{J^{(\alpha)}}{:}-
\frac{1}{2}\,{:}{\psi^{(\alpha)}}{\pd \psi^{(\alpha)}}{:}.
\end{equation}
So we see that it is the (free boson) $\times$ (free fermion) realization
of the $N=1$ superconformal algebra, the central charge of which is $c=3/2$.
From the relation~(\ref{c_cpl}) we obtain the coupling $\cpl^2=1/8$,
which means that we have to fix the level of the $so(3)$ affine algebra
to $k=4$.

The total energy-momentum field is the sum of energy-momentum fields
of the $sl(3)$ fermion system and the $so(3)$ affine algebra on level 4:
\begin{equation}
T=\frac{1}{10}\sum_{\alpha=1}^3
{:}{J^{(\alpha)}}{J^{(\alpha)}}{:}-
\frac{2}{5}\sum_{\alpha=1}^3{:}{\psi^{(\alpha)}}{\pd \psi^{(\alpha)}}{:}.
\end{equation}
The central charge is the sum of the central charge of the
$sl(3)$ fermion system ($6/5$) and the central charge of the
$so(3)$ affine vertex algebra on level 4 ($12/5$):
\begin{equation}
\CC=6/5+12/5=18/5,
\end{equation}
as one would expect.

The $U^{(\alpha)}$and the $W$ fields of the $\zz$ graded $N=1$ superconformal algebra
can be expressed in terms of $sl(n)$ fermions and affine currents
using the \ope\ relations (\ref{GG ope long}) and (\ref{TT1 term})
respectively.


\subsection{Two free bosons}

\label{Two free bosons}


This is a realization of the $p=5$ unitary model in the
series~(\ref{p series}). The coupling is $\cpl^2=1/2$, the
central charge of the $N=1$ superconformal subalgebras
is $c=1$, and the total central charge is $\CC=2$.
We take two free bosons $\phi_1$ and $\phi_2$:
\begin{equation}
\phi_i(z) \phi_j(w)=-\delta_{i,j}\, \log(z-w),
\end{equation}
and built from them the vertex operators
\begin{equation}
\Gamma_\alpha (z)=c_\alpha {:}\mathrm{e}^{\mathrm{i}(\alpha,\phi)(z)}{:}.
\end{equation}
 $\alpha$ is a vector in 2-dimensional Euclidean space,
and $({\cdot}\, ,{\cdot})$ is the standard scalar product in this space.
The factors $c_\alpha$ are the so called cocycles, satisfying
a 2-cocycle algebra, the exact definition of which is not
important here.

The 3 superconformal generators are given by
\begin{equation}
G^{(\alpha)}(z)=\frac{\Gamma_\alpha (z)+\Gamma_{-\alpha} (z)}{\sqrt{2}},
\end{equation}
where $\alpha$ is the root of the $sl(3)$ algebra normalized to
$(\alpha,\alpha)=3$.

The fields $T^{(\alpha)}(z)$ are obtained from the \ope\ (\ref{GG same ope})
and coincide with the well known energy-momentum field
for the free boson system:
\begin{equation}
T^{(\alpha)}(z)=-\frac{{:}(\alpha,\pd \phi)(\alpha,\pd \phi){:}(z)}
{2(\alpha,\alpha)}.
\end{equation}
The total energy momentum field is also the standard energy-momentum field
of the system of two free bosons:
\begin{equation}
T(z)=-\frac{1}{2}
\Big({:}\pd \phi_1 \pd \phi_1{:}(z)+
{:}\pd \phi_2 \pd \phi_2{:}(z) \Big).
\end{equation}

The fields $U^{(\alpha)}(z)$ are obtained as
\begin{equation}
U^{(\alpha)} \sim {:}(\gamma,\pd \phi)
\Big(\Gamma_{\alpha}-\Gamma_{-\alpha} \Big){:},
\end{equation}
where $\gamma$ is a vector, which is orthogonal to the root $\alpha$.

The field $W$ is given by
\begin{equation}
W=\frac{\sqrt{3}}{8} \Big(
\pd \phi_1 \pd^2 \! \phi_2- \pd \phi_2 \pd^2 \!\phi_1
\Big).
\end{equation}


\section{Discussion and speculations}

\label{Discussion}


We   constructed a new chiral algebra of parafermionic type:
the $\zz$ graded $N=1$ superconformal algebra. The full set
of \ope s is presented in Appendix~\ref{OPEs}.
The algebra has one continuous parameter:
the coupling $\cpl$, and contains three $N=1$ superconformal
subalgebras of the same central charge.
We   also discussed briefly the representation theory
of the $\zz$ graded $N=1$ superconformal algebra.
However the full description of the representation
theory remains an open problem, in particular it would be
important to understand what is the Poincare--Birkhoff--Witt basis for
the highest weight modules of the algebra.
We also obtained restrictions on the possible unitary
models of the algebra, and provided two examples
of explicit unitary realizations of the algebra.

The $\zz$ graded $N=1$ superconformal algebra is a generalization
of the $\mathbb{Z}_2$ graded $N=1$ superconformal algebra.
Higher generalizations to the case of ${\mathbb{Z}_2}^n$ grading
are possible. However in the case $n>2$ there are less dimension-$3/2$
generating fields than $2^n-1$, i.e.~not every element
(different from identity)
of the ${\mathbb{Z}_2}^n$ group has a $N=1$ superconformal generator
associated with it. The natural structure in this case is
the $A_n$ type root system. One should associate with every
pair of opposite roots (the root direction) the standard
$\mathbb{Z}_2$ graded $N=1$ superconformal algebra,
generated by $G^{(\alpha)}(z)$ and $T^{(\alpha)}(z)$,
where $\alpha$ is the root direction. There are
$n(n+1)/2$ such root directions, which is much less than
$2^n-1$ for greater $n$. Then the standard
$\mathbb{Z}_2$ graded $N=1$ superconformal algebra
corresponds to the $A_1$ root system and
the $\zz$ graded $N=1$ superconformal algebra,
described in this paper, corresponds to the $A_2$ root system.
Moreover this approach can be extended to any root system
of A-D-E type.
In fact the structure of relations between the $G^{(\alpha)}$
fields is the same as that of the \ope s of the so called
simply laced fermions defined in our previous work
(Section 7 of~\cite{Noyvert:2006qp}). The most singular term
in the \ope s will be
\begin{equation}
G^{(\alpha)}(z) G^{(\beta)}(w)=
\left\{
\begin{array}{ll}
\scriptstyle
O((z-w)^0),
& \scriptstyle
\alpha \text{ and } \beta \text{ are orthogonal,}\\
\scriptstyle
\pole{3/2}{c_{\alpha,\beta} G^{(\alpha+\beta)}(w)}
+
O((z-w)^{-1/2}),
& \scriptstyle
\alpha \text{ and } \beta \text{ are not orthogonal.}
\end{array}
\right.
\end{equation}

Again we need many more fields to close the algebra:
of conformal dimensions $5/2, 3$ and maybe of higher dimensions.
But this is a subject for a separate publication.
We want just to make a few predictions here.
Since the root system has many $A_2$ root subsystems,
the algebra has many subalgebras, which are
the $\zz$ graded $N=1$ superconformal algebras.
So we expect that there will be only one free parameter,
the coupling $\cpl$, which is connected to the
central charge of the $N=1$ superconformal subalgebras
by the same relation~(\ref{c_cpl}).
The total energy-momentum field is
\begin{equation}
T(z)=\frac{1+4 \cpl^2}{1+(3 h^\lor-2)\cpl^2}
\sum_{\alpha}
T^{(\alpha)}(z),
\end{equation}
where $h^\lor$ is the dual Coxeter number of the simply laced
algebra $\mathfrak{g}$, the root system of which is used
in the construction of our parafermionic algebra.
The total central charge is
\begin{equation}
\CC=\frac{54 \cpl^2 R_{\mathfrak{g}}}
{(1+(3 h^\lor-2)\cpl^2)(1+16\cpl^2)},
\end{equation}
where $R_{\mathfrak{g}}$ is the number of root directions of
the $\mathfrak{g}$ root system.
Substituting the values of $\cpl$ corresponding to the unitary models
from~(\ref{p series})
(like in Section~\ref{Unitary models}),
we get
two series of central charge:
\begin{equation}
\CC_p=
\frac{6 R_{\mathfrak{g}}(p^2-9)  }{\left(  p +1\right)
\left((h^\lor+2) p+3 h^\lor-10 \right)},
\qquad
\begin{aligned}
p&=3,4,5,6,\ldots\\
&\text{or}\\
p&=-3,-4,-5,-6,\ldots
\end{aligned}
\end{equation}
Unitary representations can appear only at these values
of central charge or at the limit $p \to \pm \infty$
of these two series:
\begin{equation}
\CC=\frac{6 R_{\mathfrak{g}}}
{h^\lor+2}.
\end{equation}
In the case $\mathfrak{g}=sl(n)$ this ``limit'' model is realized
by the ($sl(n)$ fermions) $\times$ ($so(n)$ affine vertex algebra on level 4),
exactly in the same way as described in section~\ref{fermions times so}.

We would like also to announce here the $N=2$ superconformal algebras
of para\-fer\-mionic type. These are also associated with the
root systems of a simple Lie algebra $\mathfrak{g}$
of the A-D-E type. But now there is a dimension-$3/2$ superconformal generator
$G^{(\alpha)}(z)$ for every root $\alpha$. The fields
$G^{(\alpha)}(z)$ and $G^{(-\alpha)}(z)$ together with dimension-1
and dimension-2 fields $J^{(\alpha)}(z)$, $T^{(\alpha)}(z)$
form the standard $N=2$ superconformal algebra.
If $\alpha+\beta$ is a root, then the \ope\ of $G^{(\alpha)}(z)$
and $G^{(\beta)}(w)$ is
\begin{equation}
G^{(\alpha)}(z)G^{(\beta)}(w)=
\pole{3/2}{c_{\alpha,\beta} G^{(\alpha+\beta)}(w)}
+
\OO{-1/2},
\end{equation}
and it is not singular if $\alpha+\beta$ is not a root.
The full field content and the \ope s
defining the algebras
are not known yet
even in the $sl(3)$ case,
they are under investigation and will be reported
in~\cite{my next}.
However we already know the minimal models of these simply laced
$N=2$ superconformal algebras. They are constructed
using the idea from
\cite{FZ:parafermions:N=2},
where the minimal models of the $sl(2)$ $N=2$ superconformal algebra
are constructed from $\mathbb{Z}_N$ parafermions and one free boson.
Our minimal models are given by
\begin{equation}
\frac{\mathfrak{g}_k}{u(1)^r} \times u(1)^r,
\end{equation}
there $k$ is the level of the affine vertex algebra $\mathfrak{g}$
of A-D-E type, and $r$ is its rank.
The first part is generated by
the Gepner parafermions~\cite{Gepner:1987sm},
and the $u(1)^r$ part is just $r$ free bosons.
The main generators are obtained as
\begin{equation}
G^{(\alpha)}(z)=\psi_\alpha(z)
\Gamma_{{\frac{\sqrt{2+k}}{\sqrt{2k}}}\alpha}(z),
\end{equation}
$\psi_\alpha$ is the parafermion corresponding
to the root $\alpha$,
the root is normalized to $(\alpha,\alpha)=2$,
the vertex operators $\Gamma$ are defined
in the section~\ref{Two free bosons}.

The formula for the total central charge of these unitary minimal
models coincide with the formula for the central charge
of the affine vertex algebra $\mathfrak{g}$:
\begin{equation}
\CC_k(\mathfrak{g})=
\frac{k\, \mathrm{dim}\mathfrak{g}}
{k+h^\lor}.
\end{equation}

The algebras described in this paper may have interesting applications
to string theory.


\subsection*{Acknowledgment}

This research was supported by a Marie Curie Intra-European
Fellowships within the 6th European Community Framework Programme.



\appendix



\section{List of \ope s}

\label{OPEs}


We list here all the algebraic relations between the
basic fields of the $\zz$ graded $N=1$ superconformal algebra.
There are 10 basic fields:
$G^{(\alpha)}, T^{(\alpha)}, U^{(\alpha)}, W,\ \alpha=1,2,3$,
of conformal dimensions $3/2$, $2$, $5/2$ and $3$ respectively.
They are primary fields with
respect to the total energy-momentum field
\begin{equation}
T=\frac{1 + 4 {\cpl}^2}{1 + 7 {\cpl}^2}
\left( T^{(1)}+T^{(2)}+T^{(3)} \right).
\end{equation}
This field $T(z)$ satisfies the Virasoro algebra with central charge
\begin{equation}
\CC=\frac{162 {\cpl}^2}{\left( 1 + 7 {\cpl}^2 \right)
\left( 1 + 16 {\cpl}^2 \right) }.
\end{equation}
The central charge of the three $N=1$ superconformal subalgebras
is expressed in terms of the coupling $\cpl$ as
\begin{equation}
c=\frac{54 \cpl^2}{(1+4 \cpl^2)(1+16\cpl^2)}.
\end{equation}

In the formulae below we use some convenient notation,
the following two-index symbol:
\begin{equation}
\sigma_{\alpha \beta}=\left\{
\begin{array}{ll}
0, & \alpha = \beta,\\
\epsilon_{\alpha \beta \gamma}, & \alpha \ne \beta \ne \gamma,
\end{array}
\right.
\qquad
\alpha, \beta, \gamma = 1,2,3,
\end{equation}
(i.e.~$\sigma_{12}=\sigma_{23}=\sigma_{31}=1$
and $\sigma_{21}=\sigma_{32}=\sigma_{13}=-1$) and
the following combination of the Virasoro fields:
\begin{equation}
\Theta
^{(\alpha)}(w)=\sum_{\gamma=1}^3 \sigma_{\alpha \gamma}
T^{(\gamma)}(w).
\end{equation}

In all the \ope s below the fields on the right hand side
of the equations are taken at point $w$.
The indices $\alpha, \beta, \gamma$ inside one equation
are all different.
There is no summation on repeated indices, unless the sum
is explicitly written.

The \ope s defining the $\zz$ graded $N=1$ superconformal
algebra read
\begin{align}
                        \label{GG same ope}
G^{(\alpha)}(z)G^{(\alpha)}(w)&=
\pole{3}{1}
+\firstpole{\frac{3}{c}T^{(\alpha)} }+
\OO{0},
\\[5pt]
T^{(\alpha)}(z)G^{(\alpha)}(w)&=
\pole{2}{\frac{3}{2}G^{(\alpha)} }
+\firstpole{\pd G^{(\alpha)} }+
\OO{0},
\\[5pt]
T^{(\alpha)}(z)T^{(\alpha)}(w)&=
\pole{4}{c/2}+
\pole{2}{2T^{(\alpha)} }+
\firstpole{\pd T^{(\alpha)} }+
\OO{0},
\\[5pt]
%
G^{(\alpha)}(z)G^{(\beta)}(w)&=
\phaseab \cpl
\left(
\frac{G^{(\gamma)} }{(z-w)^{3/2}}
+\frac{\frac{1}{2}\pd G^{(\gamma)} +
\epsilon_{\alpha \beta \gamma}U^{(\gamma)} }
{(z-w)^{1/2}}
\right)+
\OO{1/2},
\\[5pt]
T^{(\alpha)}(z)G^{(\beta)}(w)&=
\frac{9{\cpl}^2}{4\left( 1 + 4{\cpl}^2 \right) }
\left(
\pole{2}{G^{(\beta)} }
+\firstpole{\frac{2}{3}\pd G^{(\beta)} -
\sigma_{\alpha \beta}\frac{4}{3}U^{(\beta)} }
\right)+
\OO{0},
\\[5pt]
T^{(\alpha)}(z)T^{(\beta)}(w)&=
\pole{4}{\frac{c^2(1+16\cpl^2)}
{72}}+
\pole{2}{
\frac{3 {\cpl}^2}{2 \left( 1 + 4 {\cpl}^2 \right) }
\left(
2T^{(\alpha)} +2T^{(\beta)} -2T^{(\gamma)} \right)}+ \nonumber \\
&
+\firstpole{
\frac{3 {\cpl}^2}{2 \left( 1 + 4 {\cpl}^2 \right) }
\left(
\pd T^{(\alpha)} +\pd T^{(\beta)} -\pd T^{(\gamma)} \right)
+\sigma_{\alpha \beta} W }
+
\OO{0},
\\[5pt]
G^{(\alpha)}(z)U^{(\alpha)}(w)&=
\pole{2}{-\frac{3}{c}\left(
\sum_{\gamma=1}^3 \sigma_{\alpha \gamma} T^{(\gamma)}
\right)}+ \nonumber \\
&
+\firstpole{-\frac{3}{4c}\left(
\sum_{\gamma=1}^3 \sigma_{\alpha \gamma} \pd T^{(\gamma)}
\right)-\frac{27}{c^2(1+16 g^2)}W }
+\OO{0},
\\[5pt]
                                    \label{GU ope}
G^{(\alpha)}(z)U^{(\beta)}(w)&=
\frac{\phaseab}{4\cpl} \Bigg(
\pole{5/2}{-(2 + 5 {\cpl}^2)\sigma_{\alpha \beta} G^{(\gamma)} }+
\pole{3/2}{-\frac{(2 + 5 {\cpl}^2)}{6}
\sigma_{\alpha \beta} \pd G^{(\gamma)}
-\frac{2 + 17 {\cpl}^2}{3}U^{(\gamma)} }+
 \nonumber \\
&+ 
\pole{1/2}{\cpl^2 \sigma_{\alpha \beta} \pd^2 G^{(\gamma)} +
2 \cpl^2 \pd U^{(\gamma)}
-\frac{6}{c} \sigma_{\alpha \beta}
\np{T^{(\alpha)}}{G^{(\gamma)}}{0}
}
 \nonumber
\Bigg)+ \\
&+\pole{1/2}{-\frac{1}{4} \sigma_{\alpha \beta}
\np{G^{(\alpha)}}{G^{(\beta)}}{-1/2} }
+\OO{1/2},
\end{align}
\begin{align}
T^{(\alpha)}(z)U^{(\alpha)}(w)&=
\pole{2}{\frac{1 + 16 {\cpl}^2}{2 \left( 1 + 4 {\cpl}^2 \right) }
U^{(\alpha)} }
+ \nonumber \\
&+\firstpole{\frac{1 + 16 {\cpl}^2}{2 \left( 1 + 4 {\cpl}^2 \right) }
\pd U^{(\alpha)}
-\frac{1 + 16 {\cpl}^2}{12 {\cpl}^2}
\sum_{\gamma=1}^3 \sigma_{\alpha \gamma}
\np{T^{(\gamma)}}{G^{(\alpha)}}{0}
}
+\OO{0},
\\[5pt]
T^{(\alpha)}(z)U^{(\beta)}(w)&=
\frac{1}
{4 ( 1 + 4 {\cpl}^2 ) }
\Bigg(
\pole{3}{-
{3 ( 2 + 5 {\cpl}^2 ) }
\sigma_{\alpha \beta} G^{(\beta)} }+
 \nonumber
\\
&+
\pole{2}{-
{( 2 + 5 {\cpl}^2 ) }
\sigma_{\alpha \beta} \pd G^{(\beta)}
+
({4 + 19 {\cpl}^2})
U^{(\beta)} }+
 \nonumber \\
&+
\firstpole{
{3 {\cpl}^2}
  \sigma_{\alpha \beta} \pd^2 G^{(\beta)} -
{6 {\cpl}^2}
\pd U^{(\beta)}
-\frac{18}{c 
}
\sigma_{\alpha \beta} \np{T^{(\alpha)}}{G^{(\beta)}}{0}
} \Bigg)
 \nonumber
+ \\
&+\firstpole{-\frac{\phaseab}{2\cpl}\sigma_{\alpha \beta}
\np{G^{(\alpha)}}{G^{(\gamma)}}{-1/2} }
+\OO{0},
\\[5pt]
U^{(\alpha)}(z)U^{(\alpha)}(w)&=
\pole{5}{-\frac{ 2 + 5 {\cpl}^2}{4 {\cpl}^2}}
+\pole{3}{- \frac{\left( 1 + {\cpl}^2 \right)  \left( 1 + 7 {\cpl}^2 \right) }
{c \,
{\cpl}^2 \left( 1 + 4 {\cpl}^2 \right) } T +
\frac{ 2 - 13 {\cpl}^2  }{4 c \, {\cpl}^2}T^{(\alpha)} }
+ \nonumber \\
&+\pole{2}{- \frac{\left( 1 + {\cpl}^2 \right)  \left( 1 + 7 {\cpl}^2 \right) }
{2c \,
{\cpl}^2 \left( 1 + 4 {\cpl}^2 \right) } \pd T +
\frac{ 2 - 13 {\cpl}^2 }{8 c \, {\cpl}^2}\pd T^{(\alpha)} }
+\firstpole{\frac{\left( 1 + 7 {\cpl}^2 \right)  \left( 8 {\cpl}^2 -1\right) }{16 c \,
{\cpl}^2 \left( 1 + 4 {\cpl}^2 \right) }\pd^2 T
-\frac{9}{8 c}\pd^2 T^{(\alpha)} }
+ \nonumber \\
&+\firstpole{
-\frac{9}{4 c^2 {\cpl}^2}
\left(\np{T^{(\beta)}}{T^{(\gamma)}}{0}
+\np{T^{(\gamma)}}{T^{(\beta)}}{0}  \right)}
+ \nonumber \\
&+\firstpole{
-\frac{1 + 4 {\cpl}^2}{6 {\cpl}^2}
\left(\np{G^{(\beta)}}{G^{(\beta)}}{-1} +
 \np{G^{(\gamma)}}{G^{(\gamma)}}{-1}   \right)
-\frac{1}{4}\np{G^{(\alpha)}}{G^{(\alpha)}}{-1}
}
+ \nonumber \\
&+\firstpole{
-\frac{\left( 1 + 10 {\cpl}^2 \right) }{6 {\cpl}^2}
\sum_{\delta=1}^3 \sigma_{\alpha \delta}
\np{G^{(\delta)}}{U^{(\delta)}}{0} }
+\OO{0},
\end{align}
\begin{align}
U^{(\alpha)}&(z)U^{(\beta)}(w) =
\frac{\phaseab}{12\cpl} \Bigg(
\pole{7/2}{\frac{\left( 2 + 5 {\cpl}^2 \right)
\left( 2 + 17 {\cpl}^2 \right) }{4
{\cpl}^2}
 G^{(\gamma)} }+
 \nonumber
\\
&+
\pole{5/2}{\frac{\left( 2 + 5 {\cpl}^2 \right)
\left( 2 + 17 {\cpl}^2 \right) }{8
{\cpl}^2} \pd G^{(\gamma)} +
\frac{251 {\cpl}^4+ 4 {\cpl}^2-4}{12 {\cpl}^2}
\sigma_{\alpha \beta}
U^{(\gamma)} }+
 \nonumber \\
&+ 
\pole{3/2}{-\frac{4 + 19 {\cpl}^2}{4}  \pd^2 G^{(\gamma)} +
\frac{ 71 {\cpl}^4+7 {\cpl}^2 -1}{6 {\cpl}^2}
 \sigma_{\alpha \beta} \pd U^{(\gamma)} }
+
 \nonumber \\
&+
\pole{3/2}{
\frac{9 \left( 1 + {\cpl}^2 \right) }{2 \, \CC\, {\cpl}^2}
\np{T}{G^{(\gamma)}}{0}
+\frac{3 \left(11 {\cpl}^2 -1\right) }{2 \, c \, {\cpl}^2}
\np{T^{(\gamma)}}{G^{(\gamma)}}{0}
}+
 \nonumber\\
&+\pole{1/2}{
-\frac{1 + 10 {\cpl}^2}{2} \pd^3 G^{(\gamma)} +
\frac{3 \left( 1 + 4 {\cpl}^2 \right) }{2}
\sigma_{\alpha \beta} \pd^2 U^{(\gamma)}
+\frac{324}{c^2 \left( 1 + 16 {\cpl}^2 \right) }
\sigma_{\alpha \beta}\np{G^{(\gamma)}}{W}{0}
}+
 \nonumber\\
&+\pole{1/2}{\frac{81}{c^2 \left( 1 + 16 {\cpl}^2 \right) }
\pd \np{(T^{(\alpha)}+T^{(\beta)})}{G^{(\gamma)}}{0} +
\frac{9}{c}
\pd \np{(T^{(\beta)}+T^{(\gamma)})}{G^{(\gamma)}}{0}
}+
 \nonumber\\
&+\pole{1/2}{
\frac{9}{c}
 \np{(T^{(\alpha)}-T^{(\beta)})}{G^{(\gamma)}}{-1} +
\frac{18}{c} \sigma_{\alpha \beta}
 \np{T^{(\gamma)}}{U^{(\gamma)}}{0}
}
\Bigg)+ \nonumber\\
&+\pole{3/2}{-\frac{8 + 11 {\cpl}^2}{48 {\cpl}^2}
\np{G^{(\alpha)}}{G^{(\beta)}}{-1/2} }
+\pole{1/2}{
\frac{2 + 5 {\cpl}^2}{16 {\cpl}^2}
\np{G^{(\alpha)}}{G^{(\beta)}}{-3/2}
-\frac{1}{8 {\cpl}^2}
\pd \np{G^{(\alpha)}}{G^{(\beta)}}{-1/2}
}+ \nonumber\\
&+\pole{1/2}{\frac{1 + 13 {\cpl}^2}{12 {\cpl}^2}
\sigma_{\alpha \beta}
\left(\np{G^{(\alpha)}}{U^{(\beta)}}{-1/2} +
\np{U^{(\alpha)}}{G^{(\beta)}}{-1/2}  \right)
}
+\OO{1/2},
\\[5pt]
G^{(\alpha)}&(z)W(w) =
\frac{2 {\left( 1 + 4 {\cpl}^2 \right) }^2}{3 {\cpl}^2}
\Bigg(
\pole{2}{-2 \left( 1 + {\cpl}^2 \right)
U^{(\alpha)}
}+ \nonumber \\
&
+\firstpole{- \left( 1 + 10{\cpl}^2 \right)
\pd U^{(\alpha)}
+\frac{9}{c}
\sum_{\gamma=1}^3 \sigma_{\alpha \gamma}
\np{T^{(\gamma)}}{G^{(\alpha)}}{0}
}
\Bigg)+
\OO{0},
\\[5pt]
T^{(\alpha)}&(z)W(w)=
\pole{3}{\frac{6 {\cpl}^2 \left( 1 + {\cpl}^2 \right) }
{{\left( 1 + 4 {\cpl}^2 \right)}^2}
\Theta^{(\alpha)} }
+\pole{2}{
\frac{3 {\cpl}^2 \left( 1 + {\cpl}^2 \right) }
{2 {\left( 1 + 4 {\cpl}^2 \right) }^2}
\pd \Theta^{(\alpha)}
+\frac{3 c}{\CC}W
}+ \nonumber \\
&
+\firstpole{
\frac{1 + 16 {\cpl}^2}{2 \left( 1 + 4 {\cpl}^2 \right) }
\np{T^{(\alpha)}}{\Theta^{(\alpha)}}{0}
+\frac{54 {\cpl}^4}{{\left( 1 + 4 {\cpl}^2 \right) }^3}
\np{G^{(\alpha)}}{U^{(\alpha)}}{0}
}+ \nonumber \\
&
+\firstpole{-\frac{9 {\cpl}^4}{2 {\left( 1 + 4 {\cpl}^2 \right) }^2}
\pd^2 \Theta^{(\alpha)} -
\frac{3 {\cpl}^2}{1 + 4 {\cpl}^2}
\pd W
}
+\OO{0},
\end{align}

\begin{align}
U^{(\alpha)}&(z)W(w) =
\frac{1}{{\left( 1 + 4 {\cpl}^2 \right) }^2}
\Bigg(
\pole{4}{\frac{3\left( 1 + {\cpl}^2 \right)
\left( 2 + 5 {\cpl}^2 \right) }{4}
G^{(\alpha)}
}+
\pole{3}{\frac{\left( 1 + {\cpl}^2 \right)
\left( 2 + 5 {\cpl}^2 \right) }{4}
\pd G^{(\alpha)}
}+\nonumber \\
&
+\pole{2}{-\frac{15 {\cpl}^2 \left( 1 + {\cpl}^2 \right) }{8}
\pd^2 G^{(\alpha)} +
\frac{81 \left( 1 - 2 {\cpl}^2 \right) }{8\CC}
\np{T}{G^{(\alpha)}}{0} +
\frac{27 \left( 6 {\cpl}^2 -1\right) }{8 c}
\np{T^{(\alpha)}}{G^{(\alpha)}}{0}
}
+\nonumber \\
&
+\pole{2}{
-\frac{ \left( 1 + 4 {\cpl}^2 \right)
\left( 1 + 10 {\cpl}^2 \right)}{8 \cpl}
\left(
{\mathrm{e}^{-\frac{\mathrm{i}\pi}{4} \sigma_{\beta \gamma} }}
\np{G^{(\beta)}}{G^{(\gamma)}}{-1/2} +
{\mathrm{e}^{\frac{\mathrm{i}\pi}{4} \sigma_{\beta \gamma} }}
\np{G^{(\gamma)}}{G^{(\beta)}}{-1/2}
\right)
}
+\nonumber \\
&
+\firstpole{
-\frac{2 + 49 {\cpl}^2 + 128 {\cpl}^4}{32}
\pd^3 G^{(\alpha)} +
\frac{27 \left( 1 - 2 {\cpl}^2 \right) }{8\CC}
\pd \np{T}{G^{(\alpha)}}{0} +
\frac{81 {\cpl}^2}{4\CC} \np{T}{G^{(\alpha)}}{-1}
}
+\nonumber \\
&
+\firstpole{
\frac{9 \left( 14 {\cpl}^2 -1\right) }{8c}
\pd \np{T^{(\alpha)}}{G^{(\alpha)}}{0} -
\frac{81 {\cpl}^2}{4c}
\np{T^{(\alpha)}}{G^{(\alpha)}}{-1} -
\frac{27 {\cpl}^2}{2c}
\np{\Theta^{(\alpha)}}{U^{(\alpha)}}{0}
}+
\nonumber \\
&+\firstpole{-\sigma_{\beta \gamma}\frac{3 \cpl \left( 1 + 4 {\cpl}^2 \right) }{2}
\left(
{\mathrm{e}^{-\frac{\mathrm{i}\pi}{4} \sigma_{\beta \gamma} }}
\np{G^{(\beta)}}{U^{(\gamma)}}{-1/2} -
{\mathrm{e}^{\frac{\mathrm{i}\pi}{4} \sigma_{\beta \gamma} }}
\np{G^{(\gamma)}}{U^{(\beta)}}{-1/2}
\right)
}+
\nonumber \\
&+\firstpole{
\frac{3 \cpl \left( 1 + 4 {\cpl}^2 \right) }{16}
\left(
{\mathrm{e}^{-\frac{\mathrm{i}\pi}{4} \sigma_{\beta \gamma} }}
\pd \np{G^{(\beta)}}{G^{(\gamma)}}{-1/2} +
{\mathrm{e}^{\frac{\mathrm{i}\pi}{4} \sigma_{\beta \gamma} }}
\pd \np{G^{(\gamma)}}{G^{(\beta)}}{-1/2}
\right)
}
\Bigg)+
\OO{0},
\\
W&(z)W(w) =
\frac{-3 {\cpl}^2}{{\left( 1 + 4 {\cpl}^2 \right) }^3}
\Bigg(
\pole{6}{\frac{c\left( 1 + {\cpl}^2 \right)
\left( 2 + 5 {\cpl}^2 \right)}{2}}+
\pole{4}{\frac{3 c\left( 1 + {\cpl}^2 \right)
\left( 2 + 5 {\cpl}^2 \right)}{\CC} T
}
+
\pole{3}{\frac{3 c\left( 1 + {\cpl}^2 \right)
\left( 2 + 5 {\cpl}^2 \right)}{2\CC} \pd T
}+
\nonumber \\
&+
\pole{2}{\frac{9c {\left( 2 {\cpl}^2 -1 \right) }^2}{16\CC}
\pd^2 T
+\frac{81 c\left(1-  8 {\cpl}^2 \right) }{4\CC^2}
\np{T}{T}{0} +
\frac{9 \left(  20 {\cpl}^2-1 \right) }{4c}
\sum_{\alpha=1}^3 \np{T^{(\alpha)}}{T^{(\alpha)}}{0}
}+
\nonumber \\
&+
\pole{2}{
-9\cpl^2
\sum_{\alpha=1}^3 \np{G^{(\alpha)}}{G^{(\alpha)}}{-1}
}
\Bigg)+
\firstpole{\np{W}{W}{1} }+
\OO{0},
\end{align}
where $\np{W}{W}{1}=\frac{1}{2}\pd \np{W}{W}{2}-
\frac{1}{24}\pd^3 \np{W}{W}{4}$.
\vspace{8pt}

The order of fields in the \ope s above is exchanged using the following rule:
\begin{equation}
\begin{aligned}
B(z)A(w)&=A(w)B(z),\\
R^{(\alpha)}(z)S^{(\alpha)}(w)&=
-S^{(\alpha)}(w)  R^{(\alpha)}(z),\\
R^{(\alpha)}(z) S^{(\beta)}(w) (z-w)^{3/2}&=
\mathrm{i}\sigma_{\alpha,\beta}
S^{(\beta)}(w) R^{(\alpha)}(z)(w-z)^{3/2},
\quad \alpha \ne \beta,
\end{aligned}
\end{equation}
where $B$ denotes any field from the set $\{T^{(1)}, T^{(2)}, T^{(3)},W\}$,
$R$ and $S$ stand for any field from the set $\{G,U\}$,
and $A$ is any of the 10 basic fields.

The generalized Jacobi identities~(\ref{Jacobi}) are satisfied modulo the following null field condition:
\begin{equation}
27\pd W +(1 + 16 {\cpl}^2)c^2
\sum_{\alpha=1}^3 \np{G^{(\alpha)}}{U^{(\alpha)}}{0}
=0.
\end{equation}


\end{document}